\documentclass[prd,preprint,nofootinbib]{revtex4}
\usepackage{graphicx}

\begin{document}

\newcommand{\gsim}{ \mathop{}_{\textstyle \sim}^{\textstyle >} }
\newcommand{\lsim}{ \mathop{}_{\textstyle \sim}^{\textstyle <} }
\newcommand{\vev}[1]{ \left\langle {#1} \right\rangle }

\newcommand{\bear}{\begin{array}}  \newcommand{\eear}{\end{array}}
\newcommand{\bea}{\begin{eqnarray}}  \newcommand{\eea}{\end{eqnarray}}
\newcommand{\beq}{\begin{equation}}  \newcommand{\eeq}{\end{equation}}
\newcommand{\bef}{\begin{figure}}  \newcommand{\eef}{\end{figure}}
\newcommand{\bec}{\begin{center}}  \newcommand{\eec}{\end{center}}
\newcommand{\non}{\nonumber}  \newcommand{\eqn}[1]{\beq {#1}\eeq}
\newcommand{\la}{\left\langle} \newcommand{\ra}{\right\rangle}
\def\lrf#1#2{ \left(\frac{#1}{#2}\right)}
\def\lrfp#1#2#3{ \left(\frac{#1}{#2}\right)^{#3}}

%

\renewcommand{\thefootnote}{\alph{footnote}}

\renewcommand{\thefootnote}{\fnsymbol{footnote}}
\preprint{DESY 07-060}
\title{Gravitino Dark Matter from Inflaton Decay}
\renewcommand{\thefootnote}{\alph{footnote}}

\author{Fuminobu Takahashi}

\affiliation{
Deutsches Elektronen Synchrotron DESY, Notkestrasse 85,
  22603 Hamburg, Germany  
}

\begin{abstract}
We discuss a scenario that gravitinos produced non-thermally by an
inflaton decay constitute dark matter in the present universe. We find
that this scenario is realized for wide ranges of the inflaton mass
and the vacuum expectation value.  What is intriguing about this
scenario is that the gravitino dark matter can have a relatively large
free streaming length at matter-radiation equality, which can be
probed by future observation on QSO-galaxy strong lens system.
\end{abstract}


\maketitle

In spite of accumulating observational data supporting the presence of
the dark matter (DM) in our universe~\cite{Bertone:2004pz}, we have
not yet identified what DM is made of. Among many candidates proposed
thus far, the gravitino, a supersymmetric parter of the graviton, is
particularly interesting, and it has been thoroughly investigated in
connection with leptogenesis~\cite{Fukugita:1986hr} and the collider
signatures~\cite{Buchmuller:2004rq}.

The gravitinos are copiously produced by particle scattering in
thermal plasma, once the decay of the inflaton reheats the universe.
If the gravitino is the lightest supersymmetric particle (LSP), it is
stable and can be a good candidate for
DM~\cite{Moroi:1993mb,Bolz:1998ek,Bolz:2000fu,Ellis:2003dn,Steffen:2006hw}.
The gravitino abundance is directly related to the reheating
temperature, $T_R$.  In particular, for the gravitino mass $m_{3/2}$
larger than $O(10)$GeV, the required reheating temperature for the
gravitinos to be DM is so high, $T_R \gtrsim
10^9$GeV~\cite{Davidson:2002qv,Buchmuller:2004nz}, that the thermal
leptogenesis scenario may work~\cite{Buchmuller:2005eh}.

However, the detailed studies on the big bang nucleosynthesis (BBN)
revealed that the abundance and the lifetime of the
next-to-lightest supersymmetric particle (NLSP) are tightly
constrained~\cite{Pospelov:2006sc}.  This drove the above attractive
scenario into a corner, since the lifetime of NLSP tends to be longer
than the BBN bound especially for $m_{3/2}$ larger than $O(10)$GeV,
signaling the need for some changes.  Several solutions has been
proposed; e.g., a late-time entropy
production~\cite{Buchmuller:2006tt,Takayama:2007du,KT} and a theory with
R-parity violation~\cite{Takayama:2000uz,Buchmuller:2007ui}.  
Another is to abandon thermal leptogenesis and consider a 
non-thermal leptogenesis 
scenario~\cite{Asaka:1999yd, Lazarides:1993sn} instead, which requires
a lower reheating temperature, $T_R \gtrsim 10^6$GeV. Then the
gravitino can account for the observed relic density even
for $m_{3/2}$ lighter than $O(10)$GeV, making it easier
to evade the constraints from the NLSP decay.  One drawback of this
approach however is that one needs to introduce ad hoc couplings of
the inflaton with the right-handed neutrinos.

Furthermore, it has been recently pointed out that the gravitinos are
generically produced by an inflaton
decay~\cite{Kawasaki:2006gs,Asaka:2006bv,Dine:2006ii,Endo:2006tf,Endo:2006qk,Endo:2007ih,Endo:2007sz}.
Since such non-thermal gravitino production generically occurs for the
most inflation models, it is worth studying how it affects the
conventional picture on the gravitino DM scenario.  In this letter, we
pursue a possibility that the gravitinos produced by an inflaton decay
constitute a dominant component of DM. Generically, the required
reheating temperature becomes lower than without the non-thermal
production. This makes it difficult to integrate the thermal
leptogenesis scenario into this framework. As we will see later,
however, the right-handed (s)neutrinos are generated by the inflaton
decay, and their subsequent decay may generate a right amount of the
baryon asymmetry via leptogenesis for certain values of the inflaton
parameters~\cite{Endo:2006nj}.  What is particularly appealing about
this scenario is that both the gravitino DM and the non-thermal
leptogenesis can be realized without introducing any couplings ad hoc
by hand, if the inflaton parameters satisfy certain conditions.  In
addition, the produced gravitinos can have a large velocity at
matter-radiation equality, which affects the growth of density
fluctuations of DM~\footnote{ Such a DM candidate with a large
velocity and its astrophysical implication was first discussed in
Refs~.\cite{Borgani:1996ag,Lin:2000qq,Hisano:2000dz}, and intensively
studied in connection with the so-called superWIMP
mechanism~\cite{Feng:2003xh}.  }. Future observations on
e.g. QSO-galaxy strong lens system~\cite{Hisano:2006cj} may be able to
support or refute this scenario.

Let us first briefly review the recent development on the gravitino
production from the inflaton decay. There are three gravitino
production processes; (a) the gravitino pair
production~\cite{Kawasaki:2006gs,Asaka:2006bv,Dine:2006ii,Endo:2006tf};
(b) spontaneous decay at tree level~\cite{Endo:2006qk}; (c)
anomaly-induced decay at one-loop level~\cite{Endo:2007ih}.  For the
processes listed above, the gravitino production rate can be expressed as
\beq
\Gamma_{3/2} = \frac{x}{32 \pi} \lrfp{\la \phi \ra}{M_P}{2} \frac{m_\phi^3}{M_P^2},
\eeq
where $m_\phi$ is the inflaton mass, $\la \phi \ra$ a vacuum
expectation value (VEV) of the inflaton, and $M_P = 2.4 \times
10^{18}$GeV the reduced Planck mass.  Here it should be noted that
$\la \phi \ra$ is evaluated at the potential minimum after inflation.
The precise value of the numerical coefficient $x$ depends on the
production processes, possible non-renormalizable couplings in the
K\"ahler potential, and the detailed structure of the supersymmetry
(SUSY) breaking sector~\cite{Endo:2007sz}.  To be concrete, let us
assume the minimal K\"ahler potential and the dynamical SUSY breaking
(DSB)~\cite{Witten:1981nf} with a dynamical scale $\Lambda$. In the
DSB scenario, the SUSY breaking field $z$ can acquire a large mass
$m_z$, which is assumed to be roughly equal to the dynamical scale
$\Lambda \sim \sqrt{m_{3/2} M_P}$ in the following. Such a
simplification does not essentially change our arguments.  For a
low-inflation model with $m_\phi < \Lambda$, the process (a) becomes
effective, and $x = 1$.  On the other hand, for the inflaton mass
larger than $\Lambda$, the processes (b) and (c) become effective
instead. The inflaton decays into the hidden quarks in the SUSY
breaking sector via Yukawa couplings (process (b)), or into the hidden
gauge sector via anomalies (process (c)).  Since the hidden quarks and
gauge bosons (and gauginos) are energetic when they are produced, they
are expected to form jets and produce hidden hadrons through the
strong gauge interactions.  The gravitinos are likely generated by the
decays of the hidden hadrons as well as in the cascade decay processes
in jets.  We denote the averaged number of the gravitinos produced per
each jet as $N_{3/2}$.  Then $x$ is given
by~\cite{Endo:2007sz}~\footnote{If the K\"ahler potential takes a form
of the sequestered type, the spontaneous decay through Yukawa
couplings is suppressed~\cite{Endo:2006qk,Endo:2007ih}.}
\beq 
x \;=\; \frac{N_{3/2}}{8 \pi^2} \left(\frac{1}{2} N_y |Y_h^2| +
N_g \alpha_h^2 (T_g^{(h)} - T_r^{(h)})^2\right), 
\eeq
where $Y_h$ and $\alpha_h$ are the Yukawa coupling and a fine
structure constant of the hidden gauge group, respectively, $N_y$
denotes a number of the final states for the process (b), $N_g$ is a
number of the generators of the gauge group, and $T_g^{(h)}$ and
$T_r^{(h)}$ are the Dynkin indices of the adjoint representation and the
matter fields in the representation $r$.  Although $x$ depends on the
structure of the SUSY breaking sector, its typical magnitude is
$O(10^{-3} - 10^{-2})$ for $m_\phi > \Lambda$~\footnote{
Roughly, we expect $N_{3/2} = O(1-10^2)$, $N_g = O(1)$, $\alpha_h =
O(0.1)$, and $T_g^{(h)} - T_r^{(h)} = O(1)$, while $Y_h$ strongly
depends on the SUSY breaking models. Note also that the gravitino can
be produced through the Yukawa interaction in the messenger sector, if
the inflaton mass is larger than the messenger scale.
}.  
To be concrete we take
\beq
x = \left\{
\bear{cc}
1 & {\rm ~~~for~~~}m_\phi < \Lambda \\
~10^{-3} {\rm~~or~~}10^{-2}& {\rm ~~~for~~~}m_\phi > \Lambda 
\eear
\right.,
\label{eq:valuex}
\eeq
in the following. 

Using the gravitino production rate given above, we can estimate the
abundance of the gravitinos non-thermally produced by an inflaton
decay:
\bea
 Y_{3/2}^{(NT)} &=& 2 \frac{\Gamma_{3/2}}{\Gamma_{\phi}} \frac{3 T_R}{4 m_\phi},\non\\
				&\simeq & 7 \times 10^{-11}\, x \lrfp{g_*}{200}{-\frac{1}{2}} \lrfp{\la \phi \ra}{10^{15}{\rm GeV}}{2}
							\lrfp{m_\phi}{10^{12}{\rm GeV}}{2} \lrfp{T_R}{10^6{\rm GeV}}{-1},
\eea
where $g_*$ counts the relativistic degrees of freedom, and
$\Gamma_\phi$ denotes the total decay rate of the inflaton that is
related to the reheating temperature as
\beq
\Gamma_\phi \;\equiv\; \lrfp{\pi^2 g_*}{10}{\frac{1}{2}} \frac{T_R^2}{M_P}.
\eeq
Equivalently, the gravitino density parameter is
\beq
\Omega_{3/2}^{(NT)} h^2 \;\simeq\; 0.02\, x  \lrfp{g_*}{200}{-\frac{1}{2}} \lrf{m_{3/2}}{1{\rm GeV}}
							 \lrfp{\la \phi \ra}{10^{15}{\rm GeV}}{2}
							\lrfp{m_\phi}{10^{12}{\rm GeV}}{2} \lrfp{T_R}{10^6{\rm GeV}}{-1},
\eeq
where $h$ is the present Hubble parameter in units of $100$km/s/Mpc.
Note that the gravitino abundance is inversely proportional to the
reheating temperature. Due to this feature, the non-thermal gravitino
production tends to require a relatively low $T_R$ to realize the
gravitino DM scenario. Indeed, by solving $\Omega_{3/2}^{(NT)}h^2 =
0.11$~\cite{Spergel:2006hy} with respect to $T_R$, we obtain
\beq
T_R \;\simeq\;  2\times 10^5{\rm\,GeV} \, x  \lrfp{g_*}{200}{-\frac{1}{2}} \lrf{m_{3/2}}{1{\rm GeV}}
							 \lrfp{\la \phi \ra}{10^{15}{\rm GeV}}{2}
							\lrfp{m_\phi}{10^{12}{\rm GeV}}{2}.
\label{eq:neededTR}						
\eeq
So, if $T_R$ is given by the above value, the non-thermally produced
gravitino has a right abundance to become a dominant component of
DM. Generically, one has to introduce a coupling of the inflaton to
the standard-model sector with an appropriate strength, in order to
realize $T_R$ given by Eq.~(\ref{eq:neededTR}).  However, there is a
natural way to induce the reheating, and we will discuss this
possibility later.

For the non-thermally produced gravitinos to account for DM, several
conditions must be met. First, the gravitino production by thermal
scatterings should give only negligible contribution to the DM
abundance.  The abundance of the gravitinos produced by thermal
scatterings is given
by~\cite{Bolz:2000fu,Kawasaki:2004yh,Pradler:2006qh}
\begin{eqnarray}
    \label{eq:Yx-new}
    \Omega_{3/2}^{(th)} h^2 &\simeq& 
    0.14\,
    \left(\frac{m_{\tilde{g}_3}}{300 {\rm GeV}}\right)^2 
    \lrfp{m_{3/2}}{1{\rm GeV}}{-1}
    \left( \frac{T_{\rm R}}{10^{8}\ {\rm GeV}} \right),
   \end{eqnarray}
where $m_{\tilde{g}_3}$ is the gluino running mass evaluated at
$T=T_R$.  Requiring $\Omega_{3/2}^{(th)} h^2$ to be less than the observed DM
abundance, $\Omega_{DM} h^2 \simeq 0.11$, $T_R$ is bounded above:
\beq
T_R \;\lesssim\; 8 \times 10^7 {\rm\,GeV}  \left(\frac{m_{\tilde{g}_3}}{300 {\rm GeV}}\right)^{-2} 
				    \lrf{m_{3/2}}{1{\rm GeV}}.
\label{eq:const-1}				    
\eeq
This constraint is valid for $m_{3/2} \gtrsim 100{\rm\, keV}$, which is
satisfied  for the parameter space concerned as shown later.

Another constraint comes from the recent discovery that, once the
inflaton acquires a non-vanishing VEV, the inflaton decays into the
visible sector through the top Yukawa coupling~\cite{Endo:2006qk}.
Due to the presence of this decay process, $T_R$ cannot be arbitrarily
low.  Indeed, it is bounded below as
\beq
T_R \;\gtrsim\; 1.9 \times 10^3 {\rm\,GeV}\, |Y_t| 
\lrfp{g_*}{200}{-\frac{1}{4}} 
\lrf{\la \phi \ra}{10^{15}{\rm \,GeV}}
\lrfp{m_\phi}{10^{12}{\rm\,GeV}}{\frac{3}{2}},
\label{eq:low-bound-on-TR}
\eeq
where $Y_t$ is the top Yukawa coupling. The inequality is saturated if
the inflaton has no direct couplings with any other fields in
superpotential~\footnote{ Note that we assume the minimal K\"ahler
potential in the Einstein frame.  }. 

The last constraint arises from the fact that the non-thermally
produced gravitinos can have a large velocity at matter-radiation
equality, in contrast to the gravitinos produced by thermal
scatterings. This not only limits the parameter space, but also
provides a possibility that the scenario may be probed by future
observation on QSO-galaxy strong lens system. Let us estimate the
comoving free streaming length of the gravitino at matter-radiation
equality, assuming that it has an initial energy, $\epsilon\,
m_\phi/2$ when produced.  For $m_\phi < \Lambda$, we have $\epsilon =
1$ since a pair of the gravitinos is directly produced by the inflaton
decay. On the other hand, for $m_\phi > \Lambda$, multiple gravitinos
are indirectly generated by the inflaton decay, and so, its energy
tends to be smaller than $m_\phi/2$; we expect $\epsilon \;\lesssim\;
O(N_{3/2}^{-1}) = O(10^{-3} - 0.1)$.  To be concrete we will take
$\epsilon = 10^{-3}$ or $10^{-2}$ for $m_\phi > \Lambda$.  The
comoving free streaming length $\lambda_{FS}$ at matter-radiation
equality is defined by
\beq
\lambda_{FS} \;\equiv \; \int_{t_D}^{t_{\rm eq}} \frac{v_{3/2}(t)}{a(t)} dt,
\label{eq:fs}
\eeq
where $a(t)$ is the scale factor, and $t_D$ and $t_{\rm eq} (\sim 2
\times 10^{12}{\rm\, sec})$ denote the time at the inflaton decay and
at matter-radiation equality, respectively.  $v_{3/2}$ is the velocity
of the gravitino, given by
\beq
\displaystyle{v_{3/2}(t) \;=\; \frac{|{\bf p}_{3/2}|}{E_{3/2}}
 \simeq \frac{\frac{\epsilon m_\phi}{2} \lrf{a_D}{a(t)}}{\sqrt{m_{3/2}^2 
 + \frac{\epsilon^2 m_\phi^2}{4} \lrfp{a_D}{a(t)}{2}}}},
\eeq
where we have approximated $m_\phi \gg m_{3/2}$, and 
$a_D$ is the scale factor at the inflaton decay.  Integrating
(\ref{eq:fs}) yields
\bea
\lambda_{FS} &\simeq& \frac{1}{H_0 \sqrt{1 + z_{\rm eq}}} X^{-1} \sinh^{-1} X,\non\\
&\sim&0.09 {\rm\, Mpc}}\, \,\epsilon  \ln{(2X)
\lrfp{g_*}{200}{-\frac{1}{4}}    		   		
   		   \lrfp{m_{3/2}}{1{\rm GeV}}{-1} \lrf{m_\phi}{10^{12}{\rm GeV}}
                 				    \lrfp{T_R}{10^5 {\rm GeV}}{-1} 
\label{eq:lambdafs}				    
\eea
with
\bea
X &\equiv& \frac{2 m_{3/2}}{\epsilon m_\phi} \frac{a_{\rm eq}}{a_D},\non\\
   &\simeq& 8\times 10^2\, \epsilon^{-1} \lrfp{g_*}{200}{-\frac{1}{4}}    		   		
   		   \lrf{m_{3/2}}{1{\rm GeV}} \lrfp{m_\phi}{10^{12}{\rm GeV}}{-1}
                 				    \lrf{T_R}{10^5 {\rm GeV}},
\label{eq:X}				    
\eea
where $H_0$ is the Hubble parameter at present, and $z_{\rm eq}$ and
$a_{\rm eq}$ are the red-shift and the scale factor at the
matter-radiation equality. In the second equation of
(\ref{eq:lambdafs}), we have assumed $X \gg 1$ and used $H_0^{-1} \sim
4 \times 10^3 {\rm\, Mpc}$ and $z_{\rm eq} \sim 3000$.  In
Eq.~(\ref{eq:X}), we have used $a_D/a_{\rm eq} = (\Gamma_\phi \cdot
t_{\rm eq})^{-1/2}$.  The constraint from Ly-$\alpha$ clouds,
$\lambda_{FS} \lesssim 1\,$Mpc, implies $X \gtrsim 450$.  We therefore
obtain a constraint on $T_R$ as
\beq
T_R \;\gtrsim\; 5 \times 10^4\,{\rm GeV}\, \epsilon \lrfp{g_*}{200}{-\frac{1}{4}}  
			   \lrfp{m_{3/2}}{1{\rm GeV}}{-1}
		     \lrf{m_\phi}{10^{12}{\rm GeV}}.
\label{eq:const-3}
\eeq
The meaning of this constraint is clear: the reheating must occur so early 
that the velocity of the produced gravitino becomes small enough due to redshift
by the matter-radiation equality.

Thus, if the reheating temperature $T_R$ is given by
(\ref{eq:neededTR}) and satisfies the above constraints
(\ref{eq:const-1}), (\ref{eq:low-bound-on-TR}), and (\ref{eq:const-3})
in addition to the BBN constraint $T_R \gtrsim
10$\,MeV~\cite{Kawasaki:1999na,Hannestad:2004px,Ichikawa:2005vw}, the
non-thermally produced gravitinos account for DM.  As mentioned above,
one may have to add appropriate couplings of the inflaton to light
degrees of freedom, in order to realize $T_R$ given by
(\ref{eq:neededTR}).  However there is one interesting possibility
that the reheating is induced by the decay through the top Yukawa
coupling. Then the inequality (\ref{eq:low-bound-on-TR}) becomes
saturated. This is the case if there are no direct couplings of the
inflaton with any other fields in the superpotential.  The presence of
the decay process through the top Yukawa coupling not only constrains
the reheating temperature, but also provides an intriguing way to
induce the reheating. For the moment let us pursue this possibility.
From (\ref{eq:neededTR}) and (\ref{eq:low-bound-on-TR}), we obtain
\beq
 \lrf{\la \phi \ra}{10^{15}{\rm GeV}} \lrfp{m_\phi}{10^{12}{\rm GeV}}{\frac{1}{2}}
\;\simeq\;0.01\,  |Y_t| x^{-1} \lrfp{g_*}{200}{\frac{1}{4}} \lrfp{m_{3/2}}{1{\rm GeV}}{-1}.
\label{eq:rel}
\eeq
Thus, if the inflaton parameters, $m_\phi$ and $\la \phi \ra$, satisfy
the above relation (\ref{eq:rel}), the non-thermally produced
gravitino has a just right abundance to be DM. Interestingly, the free
streaming length becomes independent of the inflaton parameters and
the gravitino mass in this case. Indeed, $\lambda_{FS}$ is
approximately given by
\beq
\lambda_{FS} \;\simeq\;
1\times 10^2 {\rm\, kpc} \lrfp{g_*}{200}{-\frac{1}{4}} \lrfp{|Y_t|}{0.6}{-2}
	\lrf{\epsilon\, x}{10^{-5}},
\eeq
The Ly-$\alpha$ constraint requires  $\epsilon\, x \lesssim 10^{-4}$, 
which is naturally satisfied for a high-scale inflation 
model with $m_\phi > \Lambda$. It is intriguing that the
gravitino DM scenario points to a high-scale inflation model 
with $m_\phi > \Lambda$ and predict a relatively large free streaming length,
as long as the reheating is induced by the top Yukawa coupling.
For $x = 10^{-3} \sim 10^{-2}$ and $\epsilon = 10^{-3} \sim 10^{-1}$,
the comoving free streaming length takes a value from $10$ kpc up to
$1$ Mpc (limited by the Ly-$\alpha$ constraint).

Now let us consider the inflaton decay into the right-handed
(s)neutrinos thorough large Majorana mass terms:
\beq
W \;=\; \frac{M_i}{2}  N_i N_i,
\eeq
where $i = 1,2,3$ is the family index.  We consider the inflaton decay
into the lightest right-handed (s)neutrino $N_1$ for simplicity,
assuming that the decay into the heavier ones, $N_2$ and $N_3$, are
kinematically forbidden.  We drop the family index in the following.
The partial decay rate of the inflaton into the right-handed
(s)neutrinos is [cf.~\cite{Endo:2006qk}]
\beq
\label{eq:decay-N}
\Gamma_N \;\simeq\; \frac{1}{16 \pi} \lrfp{\la \phi \ra}{M_P}{2} 
\frac{m_\phi M^2 }{M_P^2} \sqrt{1-\frac{4M^2}{m_\phi^2}},
\eeq
where we have taken account of both the decay into the right-handed
neutrinos and that into the right-handed sneutrinos. Note that one does not
have to introduce any direct couplings of the inflaton with the right--handed
neutrinos to induce the decay. The decay proceeds as long as the inflaton
acquires a nonzero VEV.

The lepton asymmetry can be produced by the decay of the right-handed
(s)neutrinos, if $CP$ is violated in the neutrino Yukawa
matrix~\cite{Fukugita:1986hr}.  The resultant lepton asymmetry is
given by
\beq
\frac{n_L}{s} \;\simeq\; \frac{3}{2} \epsilon_1\, B_N \frac{T_R}{m_\phi},
\eeq
where $B_N \equiv \Gamma_N/\Gamma_\phi$ denotes the branching ratio
of the inflaton decay into the (s)neutrinos.  The asymmetry parameter
$\epsilon_1$ is given by~\cite{Fukugita:1986hr,Covi:1996wh}
\beq
\epsilon_1 \simeq 2.0 \times 10^{-10} \lrf{M}{10^6{\rm GeV}} 
\lrf{m_{\nu_3}}{0.05{\rm eV}} \delta_{\rm eff},
\eeq
where $m_{\nu_3}$ is the heaviest neutrino mass and $ \delta_{\rm eff}
\leq 1$ represents the effective $CP$-violating phase. The baryon
asymmetry is obtained via the sphaleron
effect:~\cite{Khlebnikov:1988sr}
\beq
\frac{n_B}{s} \;=\; -\frac{8}{23} \frac{n_L}{s}. 
\eeq
Using  the above relations, we obtain the right amount of baryon asymmetry,
\bea
\label{eq:nbs}
\frac{n_B}{s} &\simeq& 1 \times 10^{-9}\,  \lrfp{g_*}{200}{-\frac{1}{2}} 
\lrfp{M}{10^{13}{\rm GeV}}{3} 
\lrfp{\la \phi \ra}{10^{16}{\rm GeV}}{2} \lrfp{T_R}{10^6{\rm GeV}}{-1}
\lrf{m_{\nu_3}}{0.05{\rm eV}} \delta_{\rm eff},\non\\
&\simeq& 5 \times 10^{-11}  \lrfp{g_*}{200}{-\frac{1}{4}}  \lrfp{M}{10^{13}{\rm GeV}}{3} 
\lrf{\la \phi \ra}{10^{16}{\rm GeV}} \lrfp{m_\phi}{10^{14}{\rm GeV}}{-\frac{3}{2}}
\lrf{m_{\nu_3}}{0.05{\rm eV}} \delta_{\rm eff},
\eea
where we have assumed that the inequality (\ref{eq:low-bound-on-TR})
is saturated in the second equality. Note that $M$ cannot exceed
$m_\phi/2$.  Therefore, the baryon asymmetry is proportional to
positive powers of $m_\phi$, if $M$ is set to be a value that
maximizes the asymmetry.

\begin{figure}[t!]
\begin{center}
\includegraphics[width=15cm]{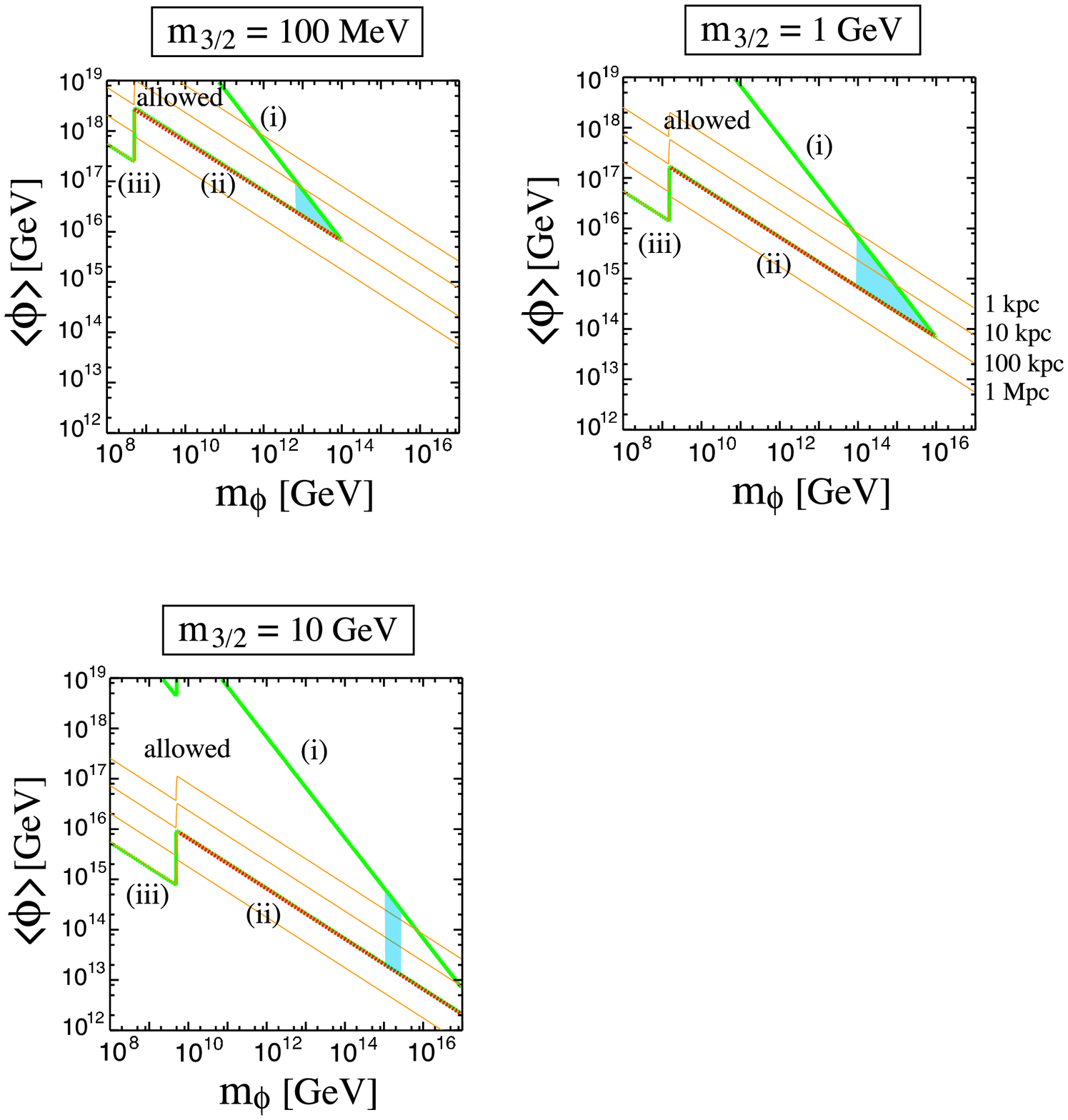}
\caption{ In the regions surrounded by the solid (green) lines, the
non-thermally produced gravitinos can account for the observed DM
density, if $T_R$ is given by (\ref{eq:neededTR}).  We have imposed
the constraints from (i) thermal production of the gravitino (see
(\ref{eq:const-1})); (ii) decay through the top Yukawa coupling (see
(\ref{eq:low-bound-on-TR})); (iii) Ly-$\alpha$ clouds (see
(\ref{eq:const-3})).  On the dotted (red) line, the reheating is
induced solely by the decay via the top Yukawa coupling and the
non-thermally produced gravitino explains DM. The thin solid (orange)
lines are the contours of the free streaming length $\lambda_{FS} =
1{\rm \,kpc}$, $10$ kpc, $100$ kpc, and $1$ Mpc, from top to bottom.
In the shaded (blue) regions the present baryon asymmetry can be
explained by the non-thermal leptogenesis.  We set $x = 10^{-3}$ and
$\epsilon = 10^{-2}$ for $m_\phi > \Lambda$, respectively. }
\label{fig:allowed}
\end{center}
\end{figure}

\begin{figure}[t!]
\begin{center}
\includegraphics[width=15cm]{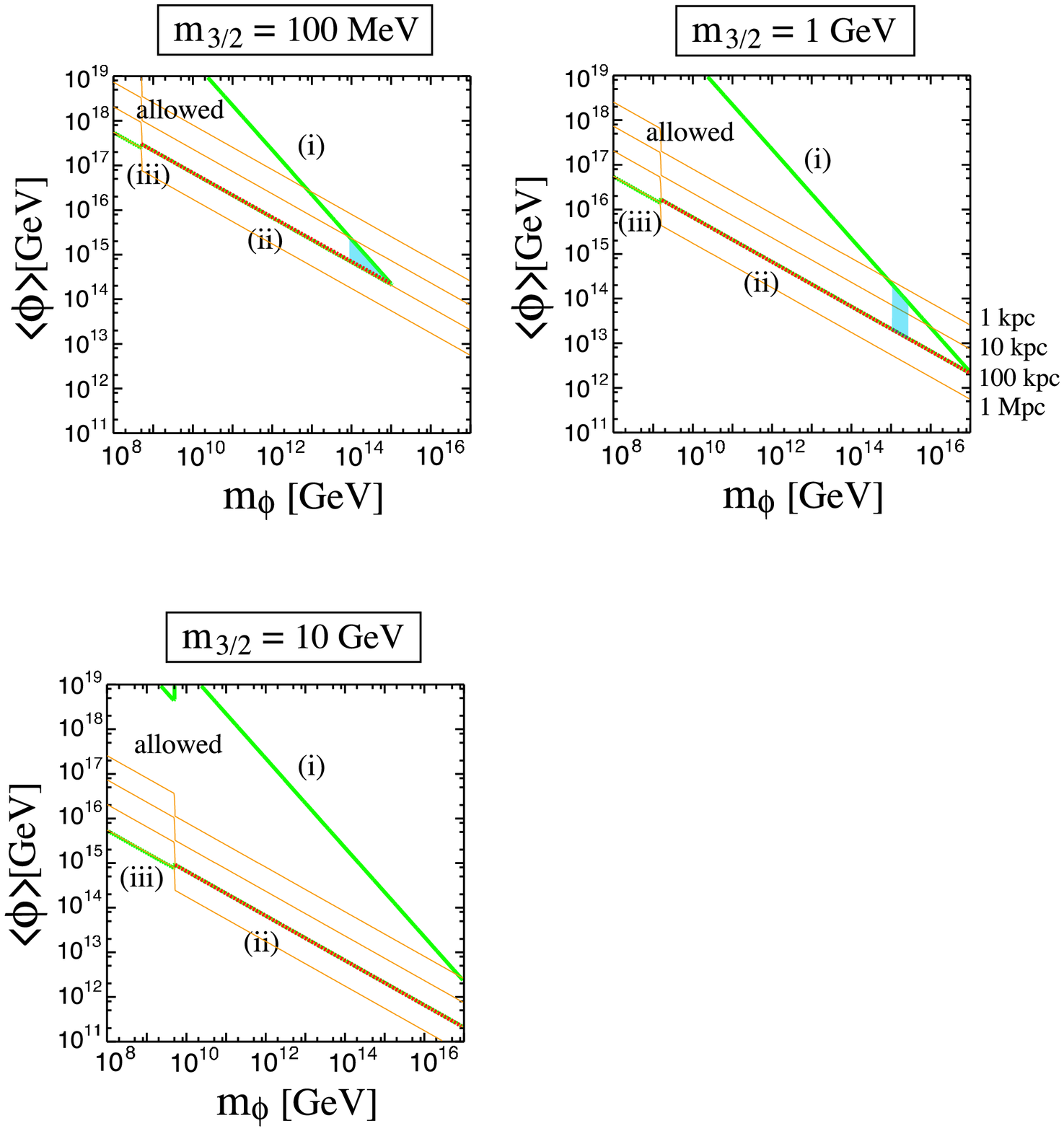}
\caption{ Same as Fig.~\ref{fig:allowed} except for $x = 10^{-2}$ and
$\epsilon = 10^{-3}$ for $m_\phi > \Lambda$.  }
\label{fig:allowed2}
\end{center}
\end{figure}

In Figs.~\ref{fig:allowed} and \ref{fig:allowed2}, we show the
parameter space where the reheating temperature (\ref{eq:neededTR})
satisfies the above constraints (\ref{eq:const-1}),
(\ref{eq:low-bound-on-TR}), and (\ref{eq:const-3}), in addition to the
BBN constraint $T_R \gtrsim 10$\,MeV. In the shaded (blue) regions,
the baryon asymmetry can be explained by the non-thermal leptogenesis
scenario discussed above, if an appropriate value of $M (\lesssim
10^{15} {\rm GeV})$ is chosen (we set $m_{\nu_3} = 0.05\,$eV and
$\delta_{\rm eff} = 1$).  We have chosen several values of the
gravitino mass: $m_{3/2} = 100{\rm \,MeV}$, $1{\rm\,GeV}$, and
$10{\rm\,GeV}$~\footnote{Note that we have not taken account of the
constraints on the NSLP decay. Therefore, the figure in the case of
$m_{3/2} = 10{\rm\,GeV}$ is valid only if the cosmological problems
associated with the NLSP are somehow avoided by e.g. introducing
R-parity violating operators with an appropriate
magnitude~\cite{Takayama:2000uz,Buchmuller:2007ui}.  }.  For smaller
$m_{3/2}$, one needs to generate more gravitinos, due to which the
allowed region shifts upward. At the same time, the constraints from
(\ref{eq:const-1}) and (\ref{eq:const-3}) become severer for smaller
$m_{3/2}$, reducing the allowed space. Thus, if $x = 10^{-3}
(10^{-2})$ for $m_\phi > \Lambda$, the gravitino mass should be larger
than $1$~MeV ($100$~keV) for the non-thermally produced gravitinos to
account for DM, since otherwise there is no allowed region for $\la
\phi \ra \lesssim M_P$.  If $x$ becomes larger for $m_\phi > \Lambda$,
the allowed region shifts downward, and a smaller value of the
gravitinos mass becomes allowed.  On the other hand, if $x$ becomes
smaller due to e.g. conformal sequestering~\cite{Endo:2007sz}, we have
more parameter space for the non-thermal leptogenesis to work
successfully.

The dotted (red) lines correspond to the special case that the
reheating is solely induced by the decay through the top Yukawa
coupling and the non-thermally produced gravitinos become DM.
Therefore the inflaton parameters on the dotted (red) lines are
particularly interesting in a sense that one does not have to
introduce any couplings ad hoc by hand; the decay spontaneous proceeds
through the top Yukawa coupling, and the gravitino has just a right
abundance to become DM.  Note that the free streaming length is
constant along the dotted (red) lines and independent of $m_{3/2}$,
$m_\phi$ and $\la \phi \ra$, as mentioned before.

 In Figs.~\ref{fig:allowed} and \ref{fig:allowed2}, we also show the
contours of the free streaming length $\lambda_{FS} = 1{\rm\, kpc}$,
$10$ kpc, $100$ kpc, and $1$ Mpc. The future submillilensing
observations can cover $\lambda_{FS} \gtrsim 2$
kpc~\cite{Hisano:2006cj}.  In particular, since the interesting case
that the reheating occurs through the top Yukawa coupling (dotted red
lines) predicts the gravitino DM with a relatively large free
streaming length ($\gtrsim 10$ kpc), it can be probed by future
observations. Such a large free streaming length may also solve the
missing satellite problem~\cite{Klypin:1999uc} and the cusp
problem~\cite{Moore:1994yx}.

From the figures, one can see that relatively broad ranges of the
inflaton mass and VEV are allowed. In particular, when combined with
the non-thermal leptogenesis scenario, we are led to a high-scale
inflation model with $m_\phi > \Lambda$.  However, studying the
parameter spaces of the representative high-scale inflation models
(such as the hybrid~\cite{Copeland:1994vg} and smooth
hybrid~\cite{Lazarides:1995vr}, and
chaotic~\cite{Kawasaki:2000yn}~\footnote{Note that the inflaton can
have a large VEV in the chaotic inflation~\footnote{More precisely,
the coefficient of the linear term in the K\"ahler potential can be
large.}, if we do not impose any discrete symmetry on the
inflaton~\cite{Endo:2006nj}. } inflation models) in detail, one finds
that only small part of the parameter space actually overlaps with the
region where the non-thermal leptogenesis works, especially if $x$
takes a value on the high side $\sim 10^{-2}$; those inflation models
tend to predict lighter $m_\phi$ and larger $\la \phi \ra$ compared to
those favored by the non-thermal leptogenesis. See
Fig.~\ref{fig:models}. Such a tension may be ameliorated if one
assumes some mechanism (e.g. the conformal sequestering) to suppress
$x$ to a smaller value.

\begin{figure}[t!]
\begin{center}
\includegraphics[width=15cm]{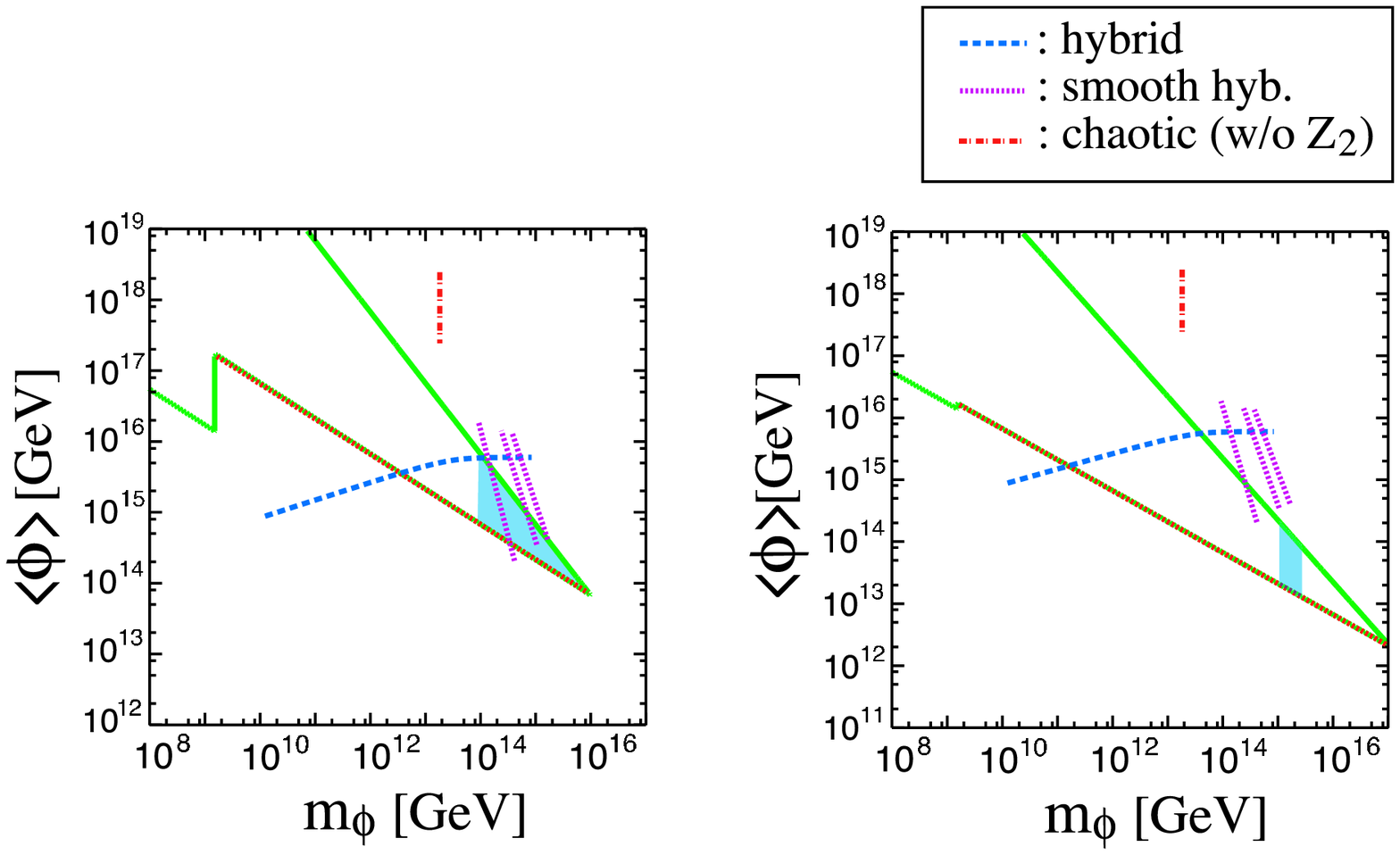}
\caption{ We show the representative high-scale inflation models;
hybrid~\cite{Copeland:1994vg} (thick long dashed (blue) line),
smooth hybrid~\cite{Lazarides:1995vr} (thick short dashed (purple) line), 
and chaotic~\cite{Kawasaki:2000yn} (long dashed dotted (red)) 
inflation models, superposed on the panels of $m_{3/2} = 1$GeV shown in 
Fig.~\ref{fig:allowed} (left) and Fig.~\ref{fig:allowed2} (right).
}
\label{fig:models}
\end{center}
\end{figure}

In summary, we have considered a scenario that the non-thermally
produced gravitinos from the inflaton decay become a dominant
component of DM.  Interestingly, if the reheating is induced solely by
the decay through the top Yukawa coupling, a high-scale inflation
model is required for the non-thermally produced gravitinos to account
for DM, and the free streaming length $\lambda_{FS}$ is predicted to
be in the range between $O(10)$\,kpc and $O(0.1)$\,Mpc, independently
of the inflaton parameters and the gravitino mass.  Such large free
streaming length may affect the growth of the density fluctuations in
DM. The suppression of the density contrast below the free streaming
scale results in the absence of the sub-halos. This feature may be
supported or refuted by future observations on the QSO-galaxy strong
lens system~\cite{Hisano:2006cj}.

\section*{Acknowledgements}
We thank W. Buchm\"uller and M. Endo for useful 
discussion and comments, and T. Takahashi for 
valuable information on the future submillilensing observations.


\end{document}